\documentclass[aps,prl,twocolumn,showpacs,superscriptaddress,groupedaddress]{revtex4}
\usepackage[pdftex]{graphicx}
\usepackage{amsmath}
\usepackage{graphicx}
\usepackage{times}
\usepackage[abs]{overpic}
\begin{document}

\title{Aggregation of superparamagnetic colloids in magnetic fields: the quest for the equilibrium state.}

\author{J.S.~Andreu}
\affiliation{Dept. de F\'isica, Universitat Aut\`onoma de Barcelona, Campus UAB, E-08193 Bellaterra, Spain.}
\affiliation{Institut de Ci\`encia de Materials de Barcelona (ICMAB-CSIC), Campus UAB, E-08193 Bellaterra, Spain.}
\author{J.~Camacho}
\affiliation{Dept. de F\'isica, Universitat Aut\`onoma de Barcelona, Campus UAB, E-08193 Bellaterra, Spain.}
\author{J.~Faraudo}
\email {jfaraudo@icmab.es}
\affiliation{Institut de Ci\`encia de Materials de Barcelona (ICMAB-CSIC), Campus UAB, E-08193 Bellaterra, Spain.}

\date{\today }
\begin{abstract}
Experimental and simulation studies of superparamagnetic colloids in strong external fields have systematically shown an irreversible aggregation process in which chains of particles steadily grow and the average size increases with time as a power-law. Here we show, by employing Langevin dynamics simulations the existence of a different aggregation behavior: aggregates form during a transient period and the system attains an equilibrium distribution of aggregate sizes. A thermodynamic self-assembly theory supports the simulation results and it also predicts that the average aggregate size in the equilibrium state depends only on a dimensionless parameter combining the volume fraction of colloids $\phi_0$ and the magnetic coupling parameter $\Gamma$. The conditions under which this new behavior can be observed are discussed.
\end{abstract}



\maketitle
Colloidal aggregation is a subject of active research for both practical (e.g. stability of many industrial products) and fundamental reasons (as a test field for statistical-mechanical theories, for example). Our interest here is in the new physics arising in the aggregation behavior of superparamagnetic colloids. These systems are a successful example of implementation of a new behavior typical of the nanoscale (superparamagnetism) in new materials with many exciting practical applications, ranging from environmental waste capture \cite{Yavuz2006} to biomedicine \cite{Krishnan2010}. Superparamagnetic materials show a large magnetic dipole in presence of external field, saturation magnetization similar to that of ferromagnetic materials but no coercitivity nor remanence at the working temperature. Superparamagnetic colloids are typically made by embedding superparamagnetic nanocrystals in a non-magnetic matrix (such as polystyrene, nanoporous silica or others) \cite{Anna}. 

Extensive experimental studies \cite{Promislow1994,Martinez,Heinrich2007,Nosaltres2008,Petersen2010} as well as computer simulations \cite{Rubio2007,Schaller2008} show that, after application of strong homogeneous and inhomogeneous magnetic fields, superparamagnetic colloids form linear aggregates. These chain-like aggregates increase in length with time, typically with a kinetic law compatible with a scale-free, power-law behavior \cite{Promislow1994,Martinez,Petersen2010}. An important property, typical of dispersions of superparamagnetic colloids, is the reversibility of chain formation: after removal of the external magnetic field, the chains rapidly disaggregate and the initial dispersion (no aggregation) is recovered \cite{Promislow1994,Martinez,Nosaltres2008,Petersen2010}. Theoretical analysis of experimental results have focused on Smoluchowsky rate equations and the appropriate kernels which reproduce the observed kinetics of chain growth under applied field \cite{Promislow1994,Martinez}.

In this work, we propose the existence of a different scenario for superparamagnetic colloids under strong external fields. Let us first note that the aggregation of superparamagnetic colloids under external field is, apparently, similar to other self-assembly processes such as micelle or gel formation. This similarity suggests that chain growth could lead, under appropriate conditions, to an equilibrium state with a constant mean chain size (and a definite distribution of chains of different sizes), as it happens in these other self-assembly processes. To the best of our knowledge, this hypothetical equilibrium state has never been reported in experiments or in simulations, suggesting that it could be difficult to realize under the conditions probed in previous studies. In this letter we will present thermodynamical arguments and brownian dynamics simulations supporting the existence of this equilibrium state under certain realistic combinations of size and saturation magnetization of the colloids. At this point, we should stress that understanding the aggregation process of superparamagnetic colloids is not only relevant from a fundamental perspective, but it has also practical importance. A paradigmatic example is the {\it fast} magnetophoretic separation process employed in biotechnological applications \cite{Nosaltres2008,Nos2009,Benelmekki2010}, which requires the formation of chains of superparamagnetic colloids.

 As in previous simulations \cite{Rubio2007,Schaller2008} we would like to consider here the minimal model describing superparamagnetic colloids: spheres of diameter $d$ with a magnetic dipole diffusing in a fluid with viscosity $\eta$. For the sake of simplicity, we assume here (as in \cite{Rubio2007}) that the magnetization of the colloids has reached saturation. This means that each colloid has a constant dipole $m_s$ (corresponding to saturation magnetization) parallel to the external applied field. This situation is also commonly found in experiments (typically at applied fields $>$0.1 T, see Ref\cite{Rubio2007,Nosaltres2008,Nos2009,Benelmekki2010}). In this situation, the magnetic effects can be described by the magnetic coupling parameter $\Gamma$ defined as the ratio between the maximum of the magnetic dipole-dipole attraction and the thermal energy:
\begin{equation}
\label{Gamma}
 \Gamma =\frac{\mu_0 m_s^2}{2\pi d^3 k_BT}.
\end{equation}
Hence, our model is characterized by two dimensionless parameters, the coupling constant $\Gamma$ and the volume fraction of colloids $\phi_0$. 

\begin{figure}
      \includegraphics[width=8.5cm]{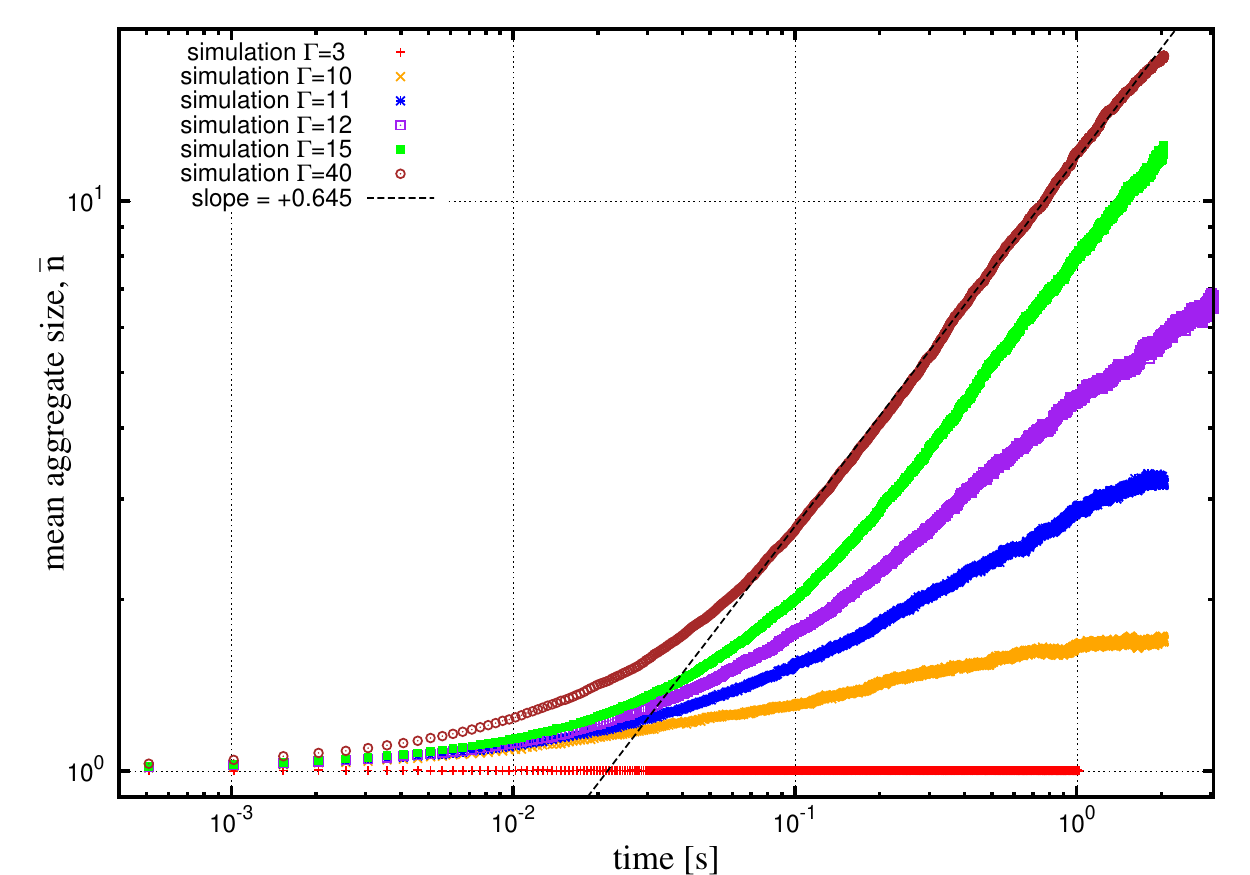}
      \caption{Time evolution of the average size of chains in simulations with 0.5 g/l concentration ($\phi_0=5.23 \times 10^{-4}$) and different magnetic coupling constants $\Gamma=3, 10, 11, 12, 15$ and 40. The dashed line is power-law fit ${\bar n}\sim t^z$ to the case $\Gamma=40$.}
      \label{Fig:Log}
 \end{figure}

\begin{figure}
      \includegraphics[width=8.5cm]{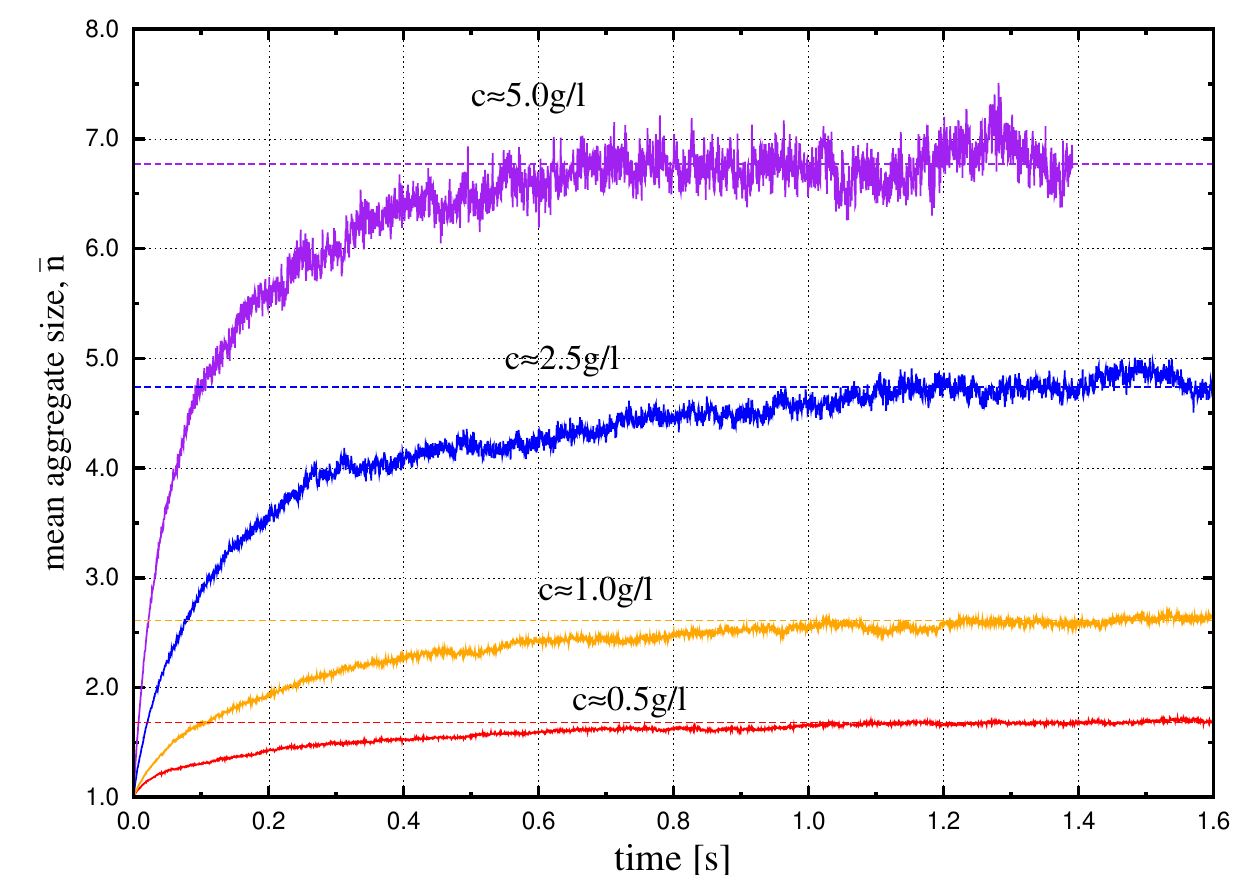}
      \caption{Time evolution of the average chain size obtained in simulations with $\Gamma=10$ and four different concentrations ($c$=0.5, 1, 2.5 and 5 g/l). Dashed lines indicate the mean aggregate size value at equilibrium.}
      \label{Fig:Nconc}
 \end{figure}

The three-dimensional simulations reported here are based on a numerical integration of the Langevin stochastic equation of motion for each colloid, as in previous works \cite{Rubio2007,Schaller2008}. In this framework, the force acting on each particle is given by the sum of a particle-particle interaction force, the viscous drag acting against each colloid and a stochastic force corresponding to the thermal noise. The particle-particle interaction potential is given by the sum of the magnetic dipole-dipole interaction and an steric, short range strong repulsion which prevents overlap between particles. We have neglected the effect of sedimentation, considering that our colloids have a density of 1 g/cm$^3$ (which is similar to that of many commercial superparamagnetic particles since it helps to avoid storage problems). All simulations were performed using the Langevin dynamics option as implemented in the 21May2008 version of the LAMMPS program \cite{LAMMPS}. The equation of motion was solved using a time step of 1 ns. Also, we employed a very large cutoff (10$d$) for the magnetic interactions in order to ensure accuracy of the results, although the resulting simulations were extremely time consuming and difficult to parallelize. Each second of simulation time requires (depending on the specific simulation) between 300-1100 hours of computer time employing 16 Itanium Montvale processors. Further technical details and movies illustrating the simulations are available in the accompanying EPAPS material\cite{EPAPS}. In all simulations, we had $N_0=8000$ colloids in the simulation box, and different concentrations were obtained by employing different system volumes $V$. The diameter of the colloids was fixed to $d=100$ nm, a value typical for small superparamagetic colloids (although this value is not essential in the sense that simulations with the same value of $\Gamma$ are expected to give equivalent results). We have considered two different simulation sets. In the first set (Fig. \ref{Fig:Log}) we considered a volume fraction of colloids $\phi_0=5.23\times 10^{-4}$ and different values of $\Gamma$. In the second simulation set (Fig. \ref{Fig:Nconc}) we have considered $\Gamma=10$ and volume fractions $\phi_0=5.23\times 10^{-4}$,$1.05\times 10^{-3}$,$2.62\times 10^{-3}$ and $5.23\times 10^{-3}$ (which corresponds to concentrations between $\sim$0.5 g/L and 5 g/L, typical of experiments).

\begin{figure}
 \begin{overpic}[width=8.5cm]{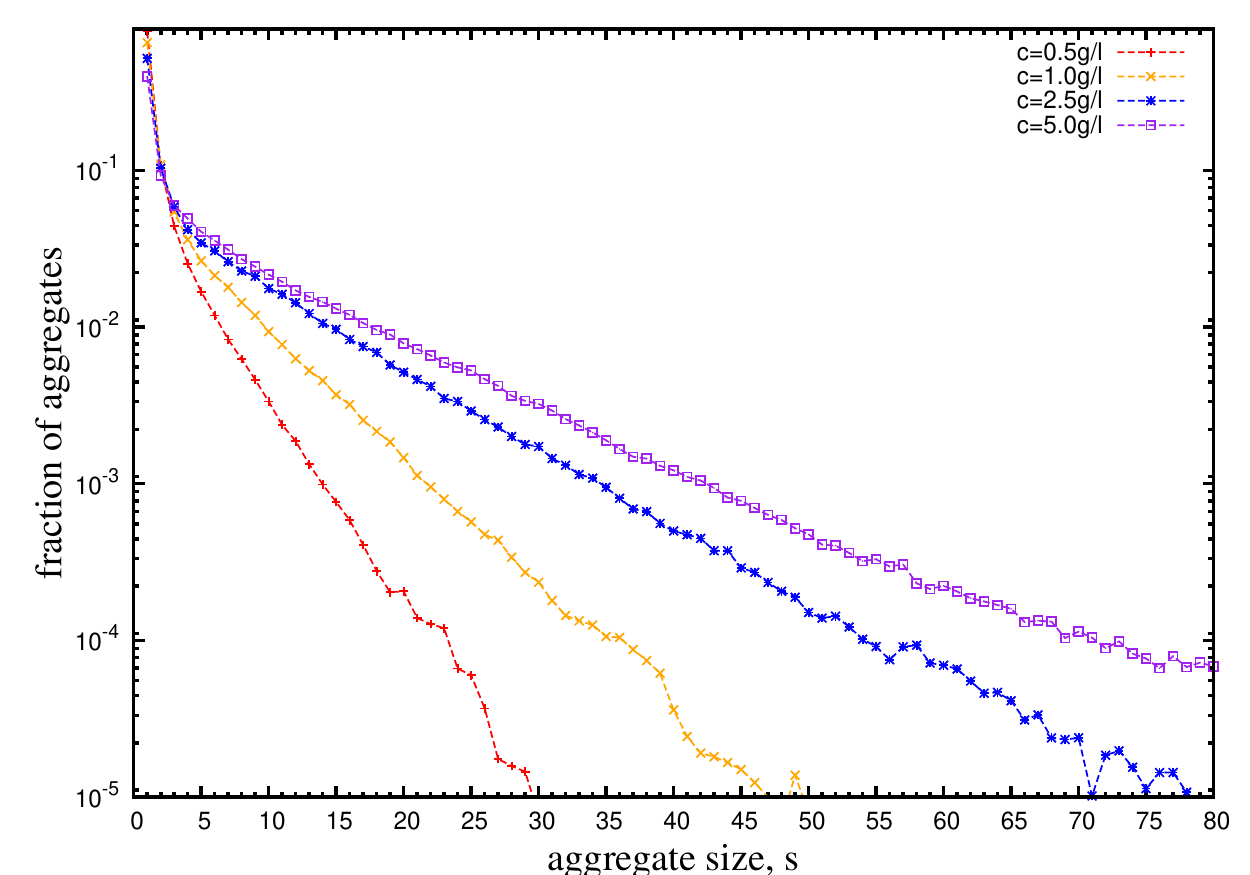}
     \put(112,77){\includegraphics[width=4.3cm]%
{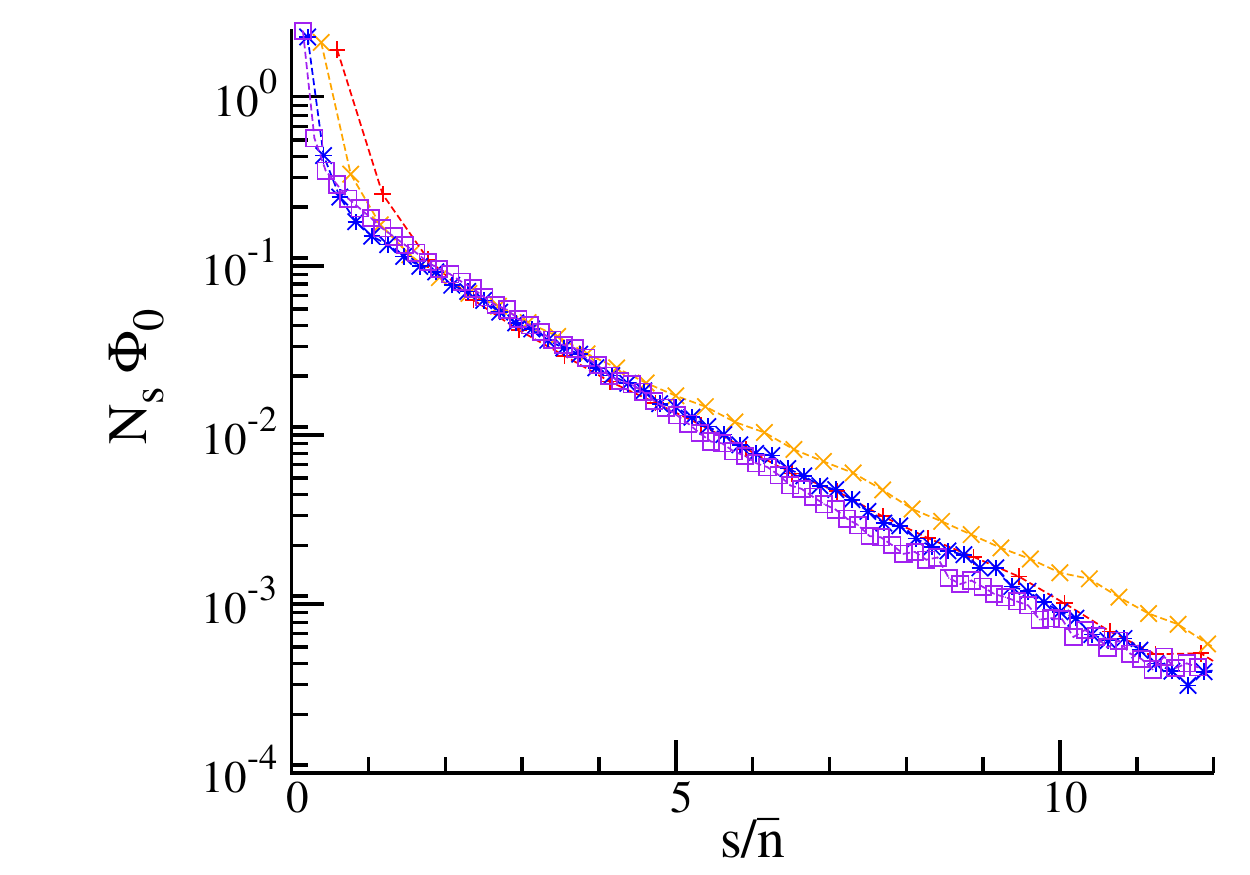}}
    \end{overpic}
      \caption{Fraction of aggregates of size $s$, $n_s/\sum_s n_s$ in the equilibrium state from simulations with $\Gamma=10$ and four different concentrations ($c$=0.5, 1, 2.5 and 5 g/l). Inset: Rescaled distribution of aggregates, $N_s \phi_0$ as a function of $s/\bar{n}$.  }
      \label{Fig:dist}
 \end{figure}
 As a check for the validity of our simulations, we looked for the typical power-law kinetic behavior observed in previous works, which consider large values of $\Gamma$ (for example in the simulations of Ref.\cite{Rubio2007} $\Gamma$ is between 100 and 3000). We have found (see Fig. \ref{Fig:Log}) that from $\Gamma$ as small as $\Gamma=15$ (and $\phi_0=5.23\times 10^{-4}$) the mean number of colloids per aggregate follows ${\bar n}\sim t^z$ with kinetic exponent $z\approx0.63$. Simulations with $\Gamma=40$ and $\phi_0=5.23\times 10^{-4}$ give a slightly larger kinetic exponent, $z\simeq0.645$. Our findings are consistent with previous works \cite{Promislow1994,Rubio2007,Martinez}, which typically report kinetic exponents in the range 0.6 - 0.7. 

Now we move towards our objective of finding an equilibrium state, with a value of ${\bar n}$ independent of time. During the time scales of our simulations, the equilibrium state was reached for simulations with $\Gamma\leq11$, as shown in Figures \ref{Fig:Log} and \ref{Fig:Nconc}. After a transient process of chain growth, we reach a time independent value of ${\bar n}$. Note that the size of the aggregates in the equilibrium state depends strongly on $\Gamma$ and $\phi_0$. At volume fraction $\phi_0=5.23 \times 10^{-4}$ of colloids, we found ${\bar n}\sim 3.2$ for $\Gamma=11$ and ${\bar n}\sim 1.7$ for  $\Gamma=10$ (see Fig. \ref{Fig:Log}). As concentration increases, the equilibrium value for ${\bar n}$ increases from $\simeq 1.7$ at 0.5 g/L to $\simeq 6.8$ at 5 g/L (see Fig. \ref{Fig:Nconc}).

Figure \ref{Fig:dist} displays the distribution of aggregate sizes in the equilibrium state (where $N_s$ is the number of aggregates of size $s$ and $n_s=N_s/V$). The main figure shows that the fraction of aggregates of size $s$, $n_s/\sum_s n_s$  decays exponentially for large $s$. The inset shows that, after an appropriate normalization, the fraction of aggregates of size $s$ for different concentrations approximately colapses in a single curve.


The obtained simulation results can be understood by considering a simple thermodynamic calculation based on the self-assembly theory \cite{Israelachvili} originally developed to describe the formation of micelles by amphiphilic molecules. In fact, the only ingredient which we need to modify in the theory is the driving force for micellization (the hydrophobic effect) which will be replaced by the magnetic interaction. Let us start by considering that the magnetic energy of a chain or aggregate made of $s$ dipoles is given by $\approx -(s-1) \epsilon_m$, i.e. the energy arises from $s-1$ bonds each one with magnetic energy $-\epsilon_m$. Although this approximation may seem rather crude, the resulting formalism captures the main features of our simulation results, as we will see. The chemical potential $\mu_s$ of a colloid which forms part of a chain of $s$ colloids is given by the ideal (entropic) term plus the interaction energy term (see Eqs.(16.1) and (16.6) in Ref.\cite{Israelachvili}):
\begin{equation}
\label{mu}
 \mu_s=\mu^{0}+ \frac{1}{s} [k_BT\ln\frac{\phi_s}{s}- (s-1)\epsilon_m ]
\end{equation}
where $\phi_s$ is the volume fraction of aggregates containing $s$ colloids, related to the number of aggregates per unit volume through $n_s \sim \phi_s/s$.  In the equilibrium state, the chemical potential $\mu_1$ of a colloid in dispersion in a non-aggregated state is equal to the chemical potential of a colloid in any of the possible chains of size $s$, so we have $\mu_1=\mu_s$. Using Eq.(\ref{mu}) we obtain ($\beta=1/k_BT$):
\begin{equation}
\label{phi}
 \phi_s=s \left[\phi_1 e^{\beta\epsilon_m}\right]^s e^{-\beta\epsilon_m}.
\end{equation}
By defining $x=\phi_1 e^{\beta\epsilon_m}$, the constraint $\phi_0=\sum_{s=1}^{\infty} \phi_s$ supplies, provided $x<1$:
\begin{equation}
\label{phi0}
 \phi_0 e^{\beta \epsilon_m}= \frac{x}{(1-x)^2}.
\end{equation}
Hence, the $x$ parameter governing the distribution of aggregate sizes in Eq. (\ref{phi}) is controlled by a single quantity defined as:
\begin{equation}
\label{N*}
N^* = \sqrt{\phi_0e^{\beta\epsilon_m}}.
\end{equation}
In the case $N^*<<1$, Eq. (\ref{phi0}) yields $\phi_1 \approx \phi_0$, i.e. no aggregation, as expected when the magnetic attraction is weak or the system is very diluted. In the opposite case, $N^*>>1$, one finds $x \approx (1- 1/N^*)$, so that Eq.(\ref{phi}) supplies:
\begin{equation}
\label{result_ns}
 n_s\sim (1-1/N^*)^s \approx e^{-s/N^*}.
\end{equation}

\begin{figure}
      \includegraphics[width=8.5cm]{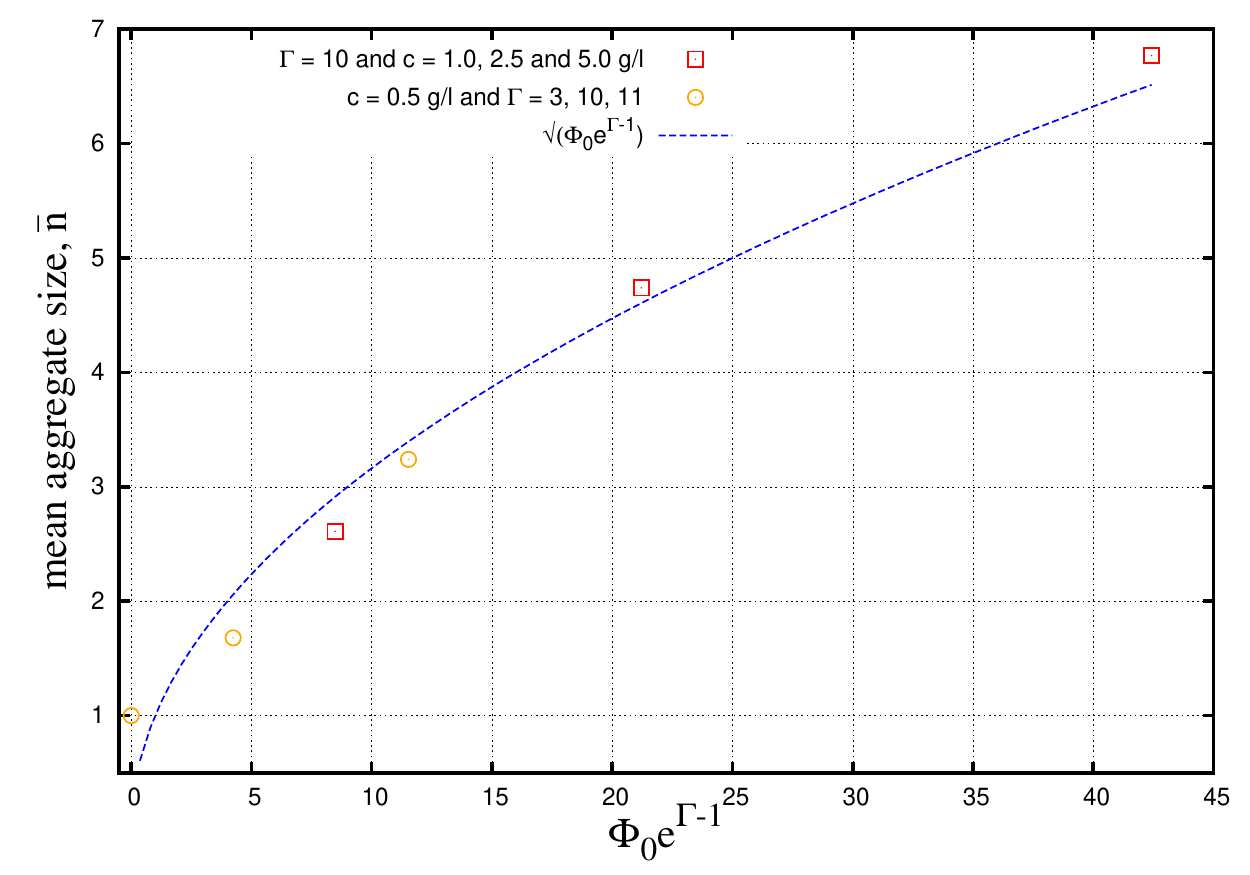} 
     
      \caption{Average number of colloids per aggregate as a function of a single parameter combining the volume fraction $\phi_0$ and the magnetic coupling $\Gamma$. The symbols correspond to simulation results for different combinations of $\phi_0$ and $\Gamma$ and the dashed line corresponds to the theoretical approximation discussed in the text.}
      \label{Fig:collapse}
 \end{figure}
In this approximation, the average size of aggregates gives ${\bar n}\approx N^*$. Also, note that an exponential decay is observed in our simulation results for large $s$ (see Fig.\ref{Fig:dist}), thereby supporting the plausibility of the simplifications introduced in the thermodynamic calculation.

In order to make more explicit predictions, we need to relate $\epsilon_m$ (and hence $N^*$) with known magnetic properties of the colloids. To this end, consider an aggregate of two colloids in contact, each one with dipole $m_s$ (corresponding to saturation magnetization) parallel to the external applied field. Their magnetic interaction energy is given by $U_m=-k_BT\Gamma (1- (3/2)\sin^2\theta)$ where $\theta$ is the angle between the magnetic field and the line joining the centers of the two colloids. This interaction is attractive for $|\theta|<\theta_0=54.7^\text{o}$, being maximum at $\theta=0$ ($\beta U_m(0)=-\Gamma$). The thermal average of this interaction over all orientations corresponding to the bonding between two colloids is given by $\beta<U_m(\theta)>=-\Gamma[1-\frac{3}{2}<\sin^2\theta>]$ where:
\begin{equation}
\label{average}
 <\sin^2\theta>=\frac{\int_{\theta=-\theta_0}^{\theta=\theta_0}d(\cos\theta) \sin^2\theta e^{-\beta U_m(\theta)}}{\int_{\theta=-\theta_0}^{\theta=\theta_0}d(\cos\theta)  e^{-\beta U_m(\theta)}}= \frac{2}{3\Gamma} +O(\frac{1}{\Gamma^2}).
\end{equation}
Hence, we have $\beta<U_m(\theta)>\simeq 1-\Gamma$. The last equality in Eq.(\ref{average}) comes from the stationary phase approximation, which in this case is extremely good. For example, for $\Gamma=10$ we obtain $\beta<U_m>=-8.96\simeq-9$ from a numerical evaluation of Eq.(\ref{average}). Even for smaller values of $\Gamma$ the approximation is quite good, for example for $\Gamma=3$ we have $\beta<U_m>=-2.071\approx-2$. Therefore, as an estimation for $\epsilon_m$ we take this thermal averaged dipole-dipole interaction, which gives $\beta\epsilon_m\approx\Gamma-1$ and then we obtain:
\begin{equation}
\label{n_gamma}
 {\bar n} \approx N^* \approx \sqrt{\phi_0 e^{(\Gamma-1)}}.
\end{equation}
The comparison shown in Fig.\ref{Fig:collapse} demonstrates that Eq.(\ref{n_gamma}) provides a fairly good approximation to the actual average size of aggregates observed in simulations. Although the number of simulation data points is small and more statistics should be desirable, it is remarkable that our simulation results are consistent with a universal behavior of the form ${\bar n}=f(\phi_0e^{(\Gamma-1)})$, as expected from the simple thermodynamic calculation. The good performance of the model is quite remarkable, given its simplicity. The result given by Eq. (\ref{n_gamma}) could be very useful in practical situations since it can be easily evaluated from characterization data (particle size, saturation magnetization and concentration) measurable in real colloidal dispersions. However, in applying Eq.(\ref{n_gamma}) in a real situation, we should keep in mind that our theoretical analysis is valid only provided that $N^*$ is much smaller than the number of colloids in the sample, $N^*<<N_0$.

 
In view of these analytical results, it is interesting to discuss again our results for the simulations which are not observed to reach an equilibrium state, at least during our simulation times (see Figure \ref{Fig:Log}). In the case of $\Gamma=40$ and $\phi_0=5.23\times 10^{-4}$, Eq.(\ref{N*}) gives $N^*\sim 6.7 \times 10^6$ and an estimation from Fig.\ref{Fig:Log} suggests an equilibration time of the order of $\sim$10 years. Clearly, in this case the reason for the no observation of the equilibrium state is that it is unphysical and it will never be observed in a real experiment. This argument also justifies why the equilibrium state is not observed in previous works, since in applications one is typically interested in large values of  $\Gamma$ in order to obtain strong magnetic effects (for example, in our previous work we employed $\Gamma\sim 10^3$ \cite{Nosaltres2008}). 

The other situation showing a power law in Fig.\ref{Fig:Log} ($\Gamma=15$ and $\phi_0=5.23\times 10^{-4}$) corresponds to a very different case. Eq.(\ref{N*}) gives $N^*$=25 and the extrapolation of the kinetics of Fig. \ref{Fig:Log} suggests an equilibration time of the order of 10 s. This calculation suggest that, in this case, we do not observe the equilibrium state due to the limitations in computational time of the simulations. However, in a real experimental situation it should be possible to observe, in this case, an initial kinetics obeying a power law followed by the more slower approach to an equilibrium state which will contain aggregates of substantial sizes. Values for $\Gamma$ around 10-15 can be easily obtained experimentally  by using superparamagnetic colloids with $d=100$ nm and saturation magnetization of $\sim$30 emu/g (see for example \cite{Anna}). Hence, the behavior reported here is readily accessible in real lab situations. 

This work is supported by the Spanish Government (grants FIS2009-13370-C02-02, PET2008-02-81-01/02 and CONSOLIDER-NANOSELECT-CSD2007-00041), the Catalan Government (grant 2009SGR164) and SEPMAG Tecnologies SL. We acknowledge computer resources and technical assistance provided by the CESGA Supercomputing center (Finisterrae Supercomputer). JF acknowledge comments by Prof. Y. Levin at the International 2nd Soft Matter Conference (Granada, Spain).
\\

\end{document}